# Spin-transfer-torque through antiferromagnetic IrMn

T. Moriyama[*,1], M. Nagata[1], K. Tanaka[1], K-J Kim[1], H. Almasi[2], W. G. Wang[2], and T. Ono[†,1]

1 Institute for Chemical Research, Kyoto University, Japan.
2 Department of Physics, The University of Arizona, USA

Abstract

Spin-transfer-torque, a transfer of angular momentum between the electron spin and the local magnetic moments, is a promising and key mechanism to control ferromagnetic materials in modern spintronic devices. However, much less attention has been paid to the same effect in antiferromagnets. For the sake of investigating how the spin current interacts with the magnetic moments in antiferromagnets, we perform spin-torque ferromagnetic resonance measurements on $Co_{20}Fe_{60}B_{20}/Ir_{25}Mn_{75}/Pt$ multilayers under a spin Hall effect of Pt. The effective magnetic damping in CoFeB is modified by the spin current injected from the Pt layer via the IrMn layer. The results indicate that the spin current interacts with IrMn magnetic moments and exerts the anti-damping torque on the magnetic moments of CoFeB through the IrMn. It is also found that the reduction of the exchange bias in the IrMn/Pt interface degrades the anti-damping torque exerted on the CoFeB layer, suggesting the transmission of the spin torque becomes less efficient as the interface exchange coupling degrades. Our work infers that the magnetic moments in IrMn can be manipulated by spin torque similarly to the one in a ferromagnetic layer.

[*]mtaka@scl.kyoto-u.ac.jp  [†]ono@scl.kyoto-u.ac.jp

Antiferromagnet (AFM) is a magnetic material which has local anisotropic magnetic moments but has no net magnetic moment as a whole. Since AFMs have no spontaneous magnetization unlike ferromagnetic materials (FMs) and the magnetic susceptibility is very small, it is generally not easy to manipulate the magnetic moments in AFMs by an external magnetic field. However, recent theoretical studies [1,2,3,4,5] suggest a possibility of manipulating the magnetic moments in AFMs by spin-transfer-torque (STT) in a similar manner to FMs[6,7].

While a number of theoretical studies have been reported on interaction between spin current and order parameters of AFM, only few experimental work have been reported so far. Wei *et al.* [8] reported that the electric current flowing in the FeMn/CoFe/Cu/CoFe spin valve can alter the exchange-bias, implying the magnetic moments in AFM is influenced by STT. Urazhdin *et al.*[9] also reported that under strong current injection in the nano-patterned spin valves the exchange bias at AFM/FM can be altered due to both the spin transfer torque and electron-magnon scattering. These previous reports[8,9] focused on the current-perpendicular-to-the-plane (CPP) measurements in the FM/AFM bilayer where the current flowing through FM is spin-polarized and interacts with the AFM magnetic moments. However, it is not clear whether or not the alternation of the exchange bias is due to the STT acting on the *bulk* AFM magnetization. It remains possible that the effect is due to the uncompensated moment at the FM/AFM interface which interacts with the STT. It is still of great interest to further experimentally investigate the interaction of spin current and AFM magnetic moments and to pursue potential applications in emerging antiferromagnetic spintronics[10,11].

The spin-torque ferromagnetic resonance (ST-FMR) measurement technique developed for spin valves[12] and magnetic tunnel junctions[13,14] and later applied to the system involved with spin Hall effect (SHE)[15] is a useful method for quantifying the spin torque exerting in the system. As above mentioned previous reports[8,9] focus only on the response of the magnetization direction under the application of the spin torque, the ST-FMR can quantify another important parameter, *magnetic damping*, which is directly connected to the interaction between the magnetic moments and the spin angular momentum carried by the spin current. In this study, we perform ST-FMR measurements on AFM/FM bilayers to experimentally investigate the transfer of the spin angular momentum into the magnetic moments in AFM. Our ST-FMR measurement enables to inject spin current directly to AFM without a help of spin-polarizing FM layer, which makes this work essentially clearer and different than the previous reports.

We prepare $Co_{20}Fe_{60}B_{20}$ 5 nm/$Ir_{25}Mn_{75}$ $t_{IrMn}$ nm/Pt 4 nm multilayers on a thermally oxidized Si substrate by magnetron sputtering. The film is patterned into 4 ~ 10 μm wide strips attached to a coplanar waveguide facilitating both the r.f. and d.c. current injection into the strip. The d.c. electric current flowing in Pt layer invokes a spin Hall effect and injects a pure spin current into the neighboring IrMn layer as shown in Fig. 1(a). ST-FMR is performed by sweeping the external magnetic field at a fixed frequency of the r.f. current. Figure 1(b) shows the measurement configuration together with our coordinate system. The positive electric current is defined when it flows along the positive *y* direction. The external positive magnetic field is applied in the sample plane and at 45º away from *x* axis. We apply nominal r.f. power up to 14 dBm and d.c. current up to 2 mA to the strip. All the measurements are performed at room temperature.

The expected rectified dc voltage $V_{dc}$ is written as[15],

$$V_{dc} = \frac{1}{4}\left(\frac{dH}{df}\right)\bigg|_{H_{ex}=H_0} \frac{dR}{d\theta} \frac{\gamma (I_{RF})^2 \sin\theta}{2\pi\sigma} (P_s S(\omega) + P_A A(\omega)), \qquad (1)$$

where $\gamma = 1.76 \times 10^{11}$ T$^{-1}$s$^{-1}$ is the gyromagnetic ratio, $H_{ex}$ is the external field, and $I_{RF}$ is r.f. current flowing in the strip. $S(\omega) = 1/((H_{ext} - H_0)^2/\sigma^2 + 1)$ and $A(\omega) = ((H_{ext} - H_0)/\sigma)/((H_{ext} - H_0)^2/\sigma^2 + 1)$ are the symmetric and the antisymmetric Lorentzian, respectively. The prefactors are $P_s = (\partial \tau_{SHE}/\partial I)(1/M_s Vol)$ and $P_A = (\partial h/\partial I)\sqrt{1 + 4\pi M_{eff}/H_{ex}}$, where $\tau_{SHE}$ is the anti-damping torque due to the SHE, $h$ is sum of the field torque and field-like torque[17,18], $M_s$ is saturation magnetization of the CoFeB, and $4\pi M_{eff}$ is the effective demagnetizing field. We expect $dR/d\theta$ to be a finite value due to anisotropic magnetoresistance (AMR) of the CoFeB. The resonant field $H_0$ and resonant frequency $\omega_0$ follow the Kittel equation as,

$$\omega_0 = \gamma\sqrt{(H_0 + H_{eb})(H_0 + H_{eb} + 4\pi M_{eff})}, \qquad (2)$$

where $H_{eb}$ is the unidirectional exchange bias field. All parameters except $\partial \tau_{SHE}/\partial I$ and $\partial h/\partial I$ in Eq. 1 are experimentally obtainable. The Lorentzian linewidth $\sigma$ is proportional to the intrinsic damping $\alpha_0$ as,

$$\sigma = \frac{\omega_0}{\gamma}\alpha_0. \qquad (3)$$

The linewidth is modified by the spin current density $(\hbar/2e)J_s$ under the spin Hall angle $\theta_{SH} = J_s/J_c$ where $J_c$ is the electric current density flowing in the spin Hall material;

$$\sigma = \frac{\omega_0}{\gamma}\left(\alpha_0 + \frac{\cos\theta}{(H_{ex} + 2\pi M_{eff})M_s t_{FM}} \frac{\hbar}{2e}\theta_{SH} J_c\right), \qquad (4)$$

where $\hbar$ is the reduced Planck constant, $e$ is the elementary charge, and $t_{FM}$ is the thickness of the magnetic layer. It should be noted that the linewidth is sensitive to the anti-damping torque not to the field-like torque.

Figure 2(a) shows typical spectra of $V_{dc}$ as a function of the external field at the different r.f frequencies in CoFeB 4nm/IrMn 23 nm/Pt 4 nm. As Eq. 1 indicates, the combination of symmetric and anti-symmetric Lorentizan can be seen around the resonant field at each r.f. frequency. We confirm that there are no hysteretic behaviors in the resonant spectra above |500| Oe (see the inset of Fig. 2(a)). It is found that the resonant fields differ in positive and negative external field as shown in Fig. 2(b), implying that there is a unidirectional exchange bias in the CoFeB layer. By fitting with Eq. 2, we extract the unidirectional anisotropy to be 47 Oe and $4\pi M_{eff}$ to be 1.4 Tesla which is consistent with the saturation magnetization of 1.4 Tesla measured by SQUID magnetometry. More precise measurements determine the unidirectional field direction to be along the positive *y* direction and the magnitude to be about 100 Oe (see supplementary information). In the case of the CoFeB 4nm/Pt 4 nm bilayer, we found $4\pi M_{eff}$ to be 1.2 Tesla due to a sizable interfacial perpendicular anisotropy at the CoFeB/Pt interface[16].

Figure 3(a) shows the ratio of $P_S$ to $P_A$ for CoFeB 4nm/IrMn $t_{IrMn}$ nm /Pt 4 nm with $t_{IrMn}$ = 0, 11, and 23 nm. Since $P_S$ and $P_A$ are proportional to $\partial \tau_{SHE}/\partial I$ and $\partial h/\partial I$, respectively, $P_S/P_A$ represents the ratio of the anti-damping torque over the sum of the field torque and field-like torque. Finite value of $P_S/P_A$ indicates that we have an anti-damping torque in the system. The monotonous increase in $P_S/P_A$ with the external field originates from the field dependence of $P_A$. It is found that $P_S/P_A$ is largest with $t_{IrMn}$ = 0 nm and becomes very small with $t_{IrMn}$ = 11 nm. It is remarkable that the $P_S/P_A$ again takes the appreciable value with thicker IrMn ($t_{IrMn}$ = 23nm). We also confirm that the sign of the $\partial \tau_{SHE}/\partial I$ derived from $P_S$ for the positive current is consistent with the spin torque direction created by the spin Hall effect in Pt layer. It would be possible to extract the magnitude of the spin torque from $P_S/P_A$ itself in a self-consistent way[15]. However, as it might mislead the final results without accurately evaluating the field-like torque arising from Rashba effect and spin-orbit torque[17,18] which may possibly be expected in our system, in the following discussion we take advantage of the linewidth analysis based on Eq. 4 rather than using $P_S/P_A$.

Figures 3(b), (c) and (d) shows the dc current $I_{cd}$ dependence of the linewidth at $f =$ 9 GHz for CoFeB 4nm/IrMn $t_{IrMn}$ nm /Pt 4 nm with $t_{IrMn}$ =0, 11, and 23 nm. As one expects from the finite value of $P_S/P_A$ discussed above, the effective damping is indeed modified by the spin current even with $t_{IrMn} \neq 0$ nm. Assuming that IrMn is not transparent for the spin polarized current (the spin diffusion length is reported to be about < 1nm[19]), the results suggest that the angular momentum of the spin current is transferred into the collection of AFM magnetic moments and it modifies the effective damping of the CoFeB layer via the exchange coupling at IrMn/CoFeB. The slope of the linear fitting $\Delta\sigma/I_{dc}$ as shown in Figs. 3 (a), (b), and (c) is plotted against the IrMn thickness in Fig. 4(a). $\Delta\sigma/I_{dc}$ is largest with $t_{IrMn} = 0$ nm and the insertion of IrMn as thin as 2 nm suddenly drops $\Delta\sigma/I_{dc}$. It is then subsided until 16 nm of IrMn. This dependence is not typically observed by the insertion of spin-diffusive paramagnetic materials like Cu[20] where injected spin current decays with a length scale of the spin relaxation of the paramagnetic materials. In this case, one would expect a monotonous decrease of $\Delta\sigma/I_{dc}$ with the thickness of the paramagnetic material insertion.

$\Delta\sigma/I_{dc}$ represents how efficient the spin current is transferred into the CoFeB layer via the IrMn layer with the form transformed from Eq. 4;

$$\Delta\sigma/I_{dc} = \frac{\omega_0}{\gamma} \frac{\cos\theta}{(H_{ex} + 2\pi M_{eff})M_s t_{FM} w} \frac{\hbar}{2e} \left(\beta \frac{\theta_{SH,Pt} r_{Pt}}{t_{Pt}}\right), \quad (5)$$

where $w$ is the width of the strip, $r_{Pt}$ is the shunt current ratio in the Pt layer, and $t_{Pt}$ is the thickness, and $\beta$ ($0 \leq \beta \leq 1$) is the spin-transfer efficiency in the IrMn layer. $\beta = 1$ when the all injected spin current exerts a spin torque on the CoFeB layer through the IrMn layer and $\beta = 0$ when it is dissipated within the IrMn layer before reaching to the CoFeB layer. To accurately evaluate $\Delta\sigma/I_{dc}$, we separately measure the resistivity of each layer to be 1.4 x $10^{-6}$, 1.9 x $10^{-6}$, and 3.0 x $10^{-7}$ $\Omega$ m for CoFeB, IrMn, and Pt, respectively, and then calculate the shunt current ratio of the Pt layer. We estimated the spin Hall angle for the Pt layer to be 0.09 ± 0.01 by a CoFeB 4nm/Pt 4nm bilayer. We also characterized the spin Hall angle for the IrMn layer and concluded that the effect is negligible (See the supplementary information). For the sample with $t_{IrMn} \neq 0$ nm, the shunt current ratio in Pt decreases with increasing the IrMn thickness.

The dotted curve in Fig. 4 is the calculated $\Delta\sigma/I_{dc}$ based on Eq. (5) by assuming $\beta = 1$. The calculated $\Delta\sigma/I_{dc}$ decreases with increasing IrMn thickness because of the decrease in the shunt current ratio in Pt by the current shunting in the IrMn layer. In other words, this theoretical curve represents the spin torque which the CoFeB layer receives based on the current flowing in the Pt layer as if there exist no IrMn insertion layer which may dissipate the spin angular momentum. The $\beta = 1$ curve does not reproduce the measured $\Delta\sigma/I_{dc}$ until 16 nm. Our results surprisingly indicate $\beta \sim 1$ for IrMn thicker than 16 nm, implying that the spin torque is interacting with the CoFeB magnetization through the IrMn layer. We speculate that this peculiar spin transfer torque through the antiferromagnet may be attributed to the angular momentum transfer mediated by antiferromagnetic spin fluctuations[21]. First, the injected spin current interacts with the magnetic moment at the Pt/IrMn interface and induces magnetic excitations. The magnetic excitation then propagates and transfers the spin angular momentum into the CoFeB layer. The discrepancy between the measured $\Delta\sigma/I_{dc}$ and the calculation in the thinner IrMn regime then indicates the breakdown of this hypothesis and can be explained by the following mechanisms.

The important factors are the exchange coupling at the IrMn/CoFeB interface and the IrMn thickness dependence of the blocking temperature[22,23,24]. Even if the magnetic excitations are transmitted from the Pt/IrMn interface, absence of the magnetic coupling at the IrMn/CoFeB interface cannot effectively transfer the angular momentum to the CoFeB layer or influence the linewidth. Our results clearly show that the strength of the exchange bias as a function of the IrMn thickness coincides with the trend of $\Delta\sigma/I_{dc}$ (see Fig. 4(a) and (b)). Namely, small exchange bias yields small β. It is also possible that the small β in thinner IrMn layer is due to the Néel temperature degradation[25] at which the angular momentum transfer by the antiferromagnetic excitation becomes inefficient.

In addition, we estimated the IrMn thickness dependence of the intrinsic damping constant in absence of the spin current as shown in Fig.4 (c). The intrinsic damping also has a similar trend to the exchange bias as well as $\Delta\sigma/I_{dc}$. Here, the intrinsic damping is estimated from the slope of the linear fitting to the frequency dependence of the linewidth (examples are shown in the inset of Fig. 4) so that the extrinsic damping contributions such as the two-magnon scattering[26] due to magnetic inhomogeneity induced at the AFM/FM interface[27] is successfully ruled out. The increase of the intrinsic damping coinciding with the emergence of the exchange bias strongly suggests that the dynamics of the FM moment is nonlocally influenced by the AFM[28]. Namely, the angular momentum of the precessing FM is dissipated into the AFM through the exchange coupled AFM/FM interface. This is another evidence that there is a channel for the angular momentum flow through the magnetic coupling at AFM/FM. Thus, all the observations support our scenario that the spin torque is mediated by the antiferromagnetic excitation through the IrMn layer.

In summary, we investigate the interaction between spin current and magnetic moments in the antiferromagnetic IrMn in the structure of CeFeB 4nm/IrMn $t_{IrMn}$ nm/Pt 4 nm by using ST-FMR technique. We find that the linewidth, that is proportional to the *effective damping*, changes as a function of the spin current even with $t_{IrMn} \neq 0$ nm. The results indicate that the spin current is transferred and exerts a spin torque on the CoFeB layer through the IrMn layer. We speculate the spin current interacts with IrMn magnetic moments and exerts the anti-damping torque on the magnetic moments of $Co_{20}Fe_{60}B_{20}$ through the IrMn. The decrease of the spin-transfer efficiency $\beta$ in the IrMn thickness < 16nm is explained by the lack of the interfacial exchange coupling. Our results manifest that a remarkable interaction between spin current and the magnetization in the antiferromagnet as theories have predicted[1,2,3,4,5]. Our work infers that the magnetic moments of the antiferromagnetic IrMn can be manipulated by the spin transfer torque similarly to the ferromagnetic materials, raising those relatively abandoned materials to an emerging antiferromagnetic spintronics[11].

**Acknowledgements**
We would like to thank Dr. So Takei and Prof. Yaroslav Tserkovnyak for fruitful discussions. This work was partly supported by Grants-in-Aid for Scientific Research (S), Grant-in-Aid for Young Scientists (B) from Japan Society for the Promotion of Science and by US NSF (ECCS-1310338).

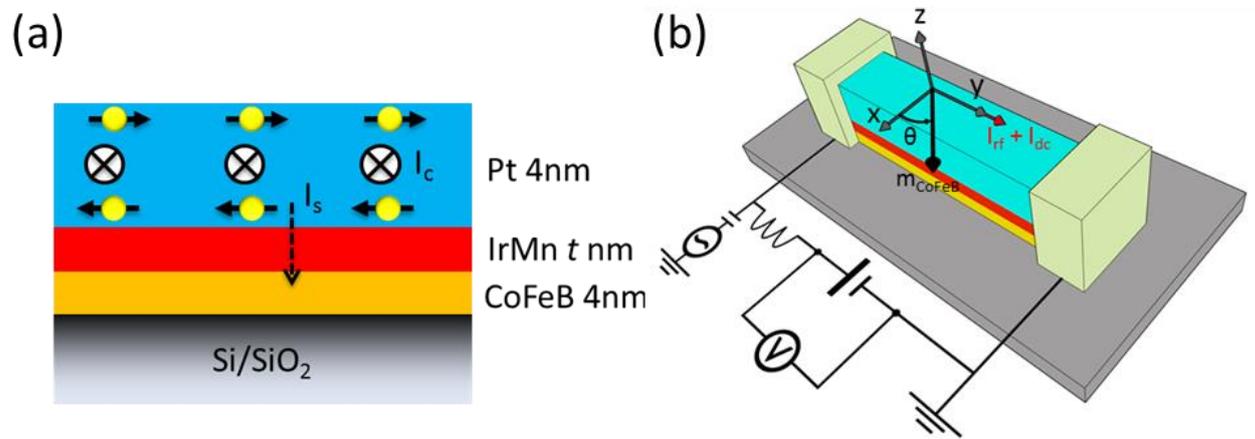

Figure 1 Moriyama et al.

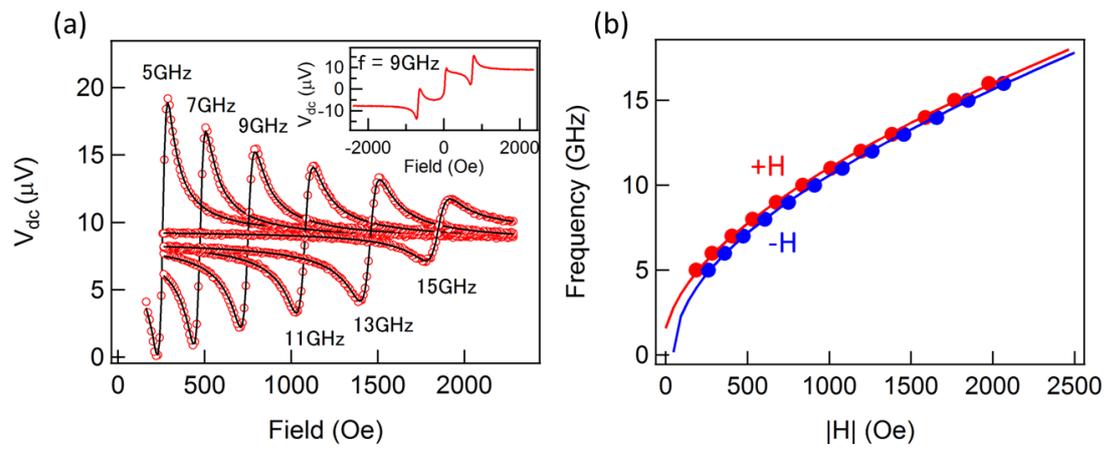

Figure 2 Moriyama et al.

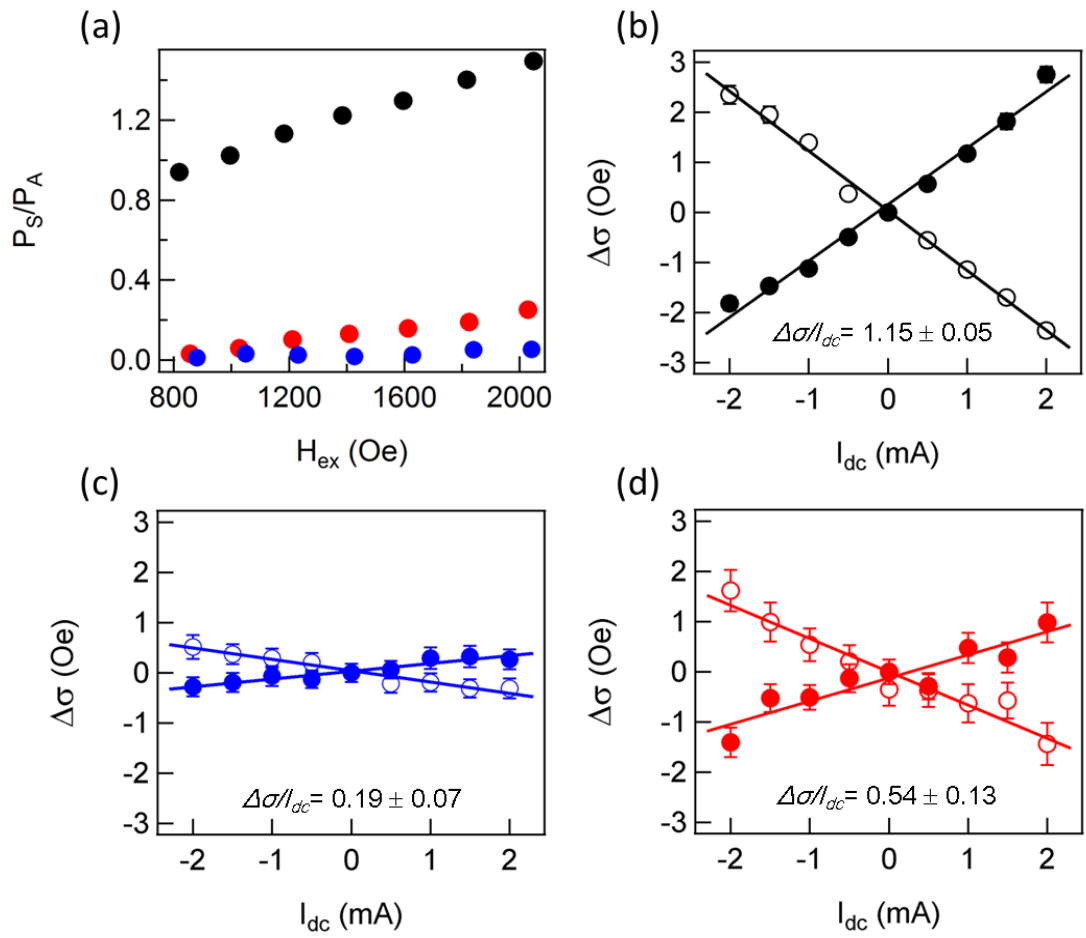

Figure 3 Moriyama et al.

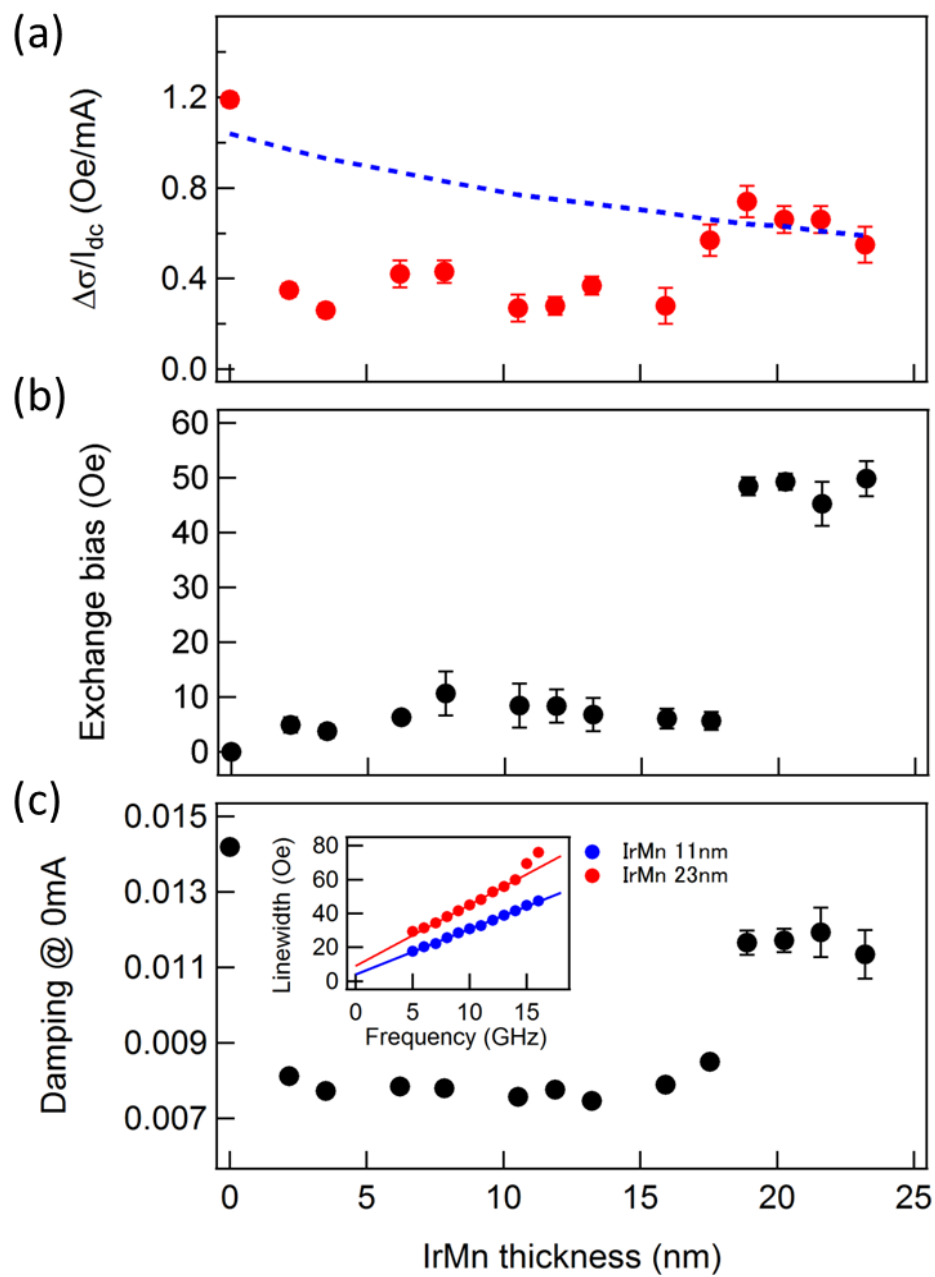

Figure 4 Moriyama et al.

FIGURE CAPTIONS:

FIG. 1 Schematic illustrations of the sample and measurement setup. (a) The sample cross section of CoFeB/IrMn/Pt deposited on a Si/SiO$_x$ substrate. The electric current flowing laterally in the Pt layer invokes the spin Hall effect and injects a spin current into the neighboring IrMn layer. (b) The sample is patterned into 4 ~ 10μm-wide strip and is connected to coplanar waveguide so that ST-FMR measurement can be performed.

FIG. 2 (a) $V_{dc}$ spectra from the ST-FMR measurements on CoFeB 4nm/IrMn 23 nm/Pt 4 nm with $I_{dc}$ = 0 mA. The r.f. frequency is varied up to 15 GHz. The black curves are fitting with Eq. 1. The inset shows the full $V_{dc}$ spectrum at 9 GHz with back and forth scanning between -2400 Oe and 2400 Oe. (b) The resonant frequency as a function of the positive (red) and negative field (blue).

FIG. 3 (a) $P_S/P_A$ as a function of $H_{ex}$ for $t_{IrMn}$ = 0 nm (black), 11 nm (blue), 23nm (red). Change in linewidth $\varDelta\sigma$ as a function of $I_{dc}$ for (b) $t_{IrMn}$ = 0 nm, (c) $t_{IrMn}$ = 11 nm, (d) $t_{IrMn}$ = 23 nm at the resonant frequency of 9 GHz. The open and solid circles are respectively for the positive field and negative field. Fitted slope is $\varDelta\sigma/I_{dc}$ = 1.19 ± 0.04 (Oe/mA) in the positive field and $\varDelta\sigma/I_{dc}$ = 1.11 ± 0.03 (Oe/mA) in the negative field for $t_{IrMn}$ = 0 nm. $\varDelta\sigma/I_{dc}$ = 0.23 ± 0.05 (Oe/mA) in the positive field and $\varDelta\sigma/I_{dc}$ = 0.15 ± 0.06 (Oe/mA) in the negative field for $t_{IrMn}$ = 11 nm. $\varDelta\sigma/I_{dc}$ = 0.44 ± 0.08 (Oe/mA) in the positive field and $\varDelta\sigma/I_{dc}$ = 0.64 ± 0.1 (Oe/mA) in the negative field for $t_{IrMn}$ = 23 nm.

FIG. 4 (a) $\varDelta\sigma/I_{dc}$ as a function of IrMn thickness. The dotted blue curve is the calculated

$\Delta\sigma/I_{dc}$ based on Eq. 4 with $\beta = 1$. (b) Exchange bias field as a function of IrMn thickness. (c) The intrinsic damping at $I_{dc} = 0$ mA as a function of IrMn thickness. The inset shows examples of the linear fitting on the frequency dependence of the linewidth for estimating the intrinsic damping[26].

# Spin-transfer-torque through antiferromagnetic IrMn

T. Moriyama[1], M. Nagata[1], K. Tanaka[1], K-J Kim[1], H. Almasi[2], W. Wang[2], and T. Ono[1]

1 Institute for Chemical Research, Kyoto University, Japan.
2 Department of Physics, The University of Arizona, USA

## Supplementary information

1. Determination of the exchange bias in Pt/IrMn/CoFeB layers

   To determine the exchange bias field $H_{eb}$, we performed anisotropic magnetoresistance (AMR) measurements in a rotating magnetic field[1]. Figure S1 shows the AMR as a function of the field angle $\theta$ with the field strength of $H_{ex} = 400$ Oe (see the definition for Fig. 1 in the main text.). When the crystalline anisotropy and the in-plane shape anisotropy are negligible comparing with the external field, the AMR curve is given as[1]

   $$\rho = \rho_0 + \Delta\rho \cos\left(\tan^{-1}\left(\frac{\cos\theta + (H_{eb}/H_{ex})\cos\theta_{eb}}{\sin\theta + (H_{eb}/H_{ex})\sin\theta_{eb}}\right)\right), \qquad (S1)$$

   where $\rho_0$ is the resistivity at $\theta = 90°$ and $\theta_{eb}$ is the direction of the exchange bias. The fitting with Eq. S1 yields $\theta_{eb} = 90°$ and $H_{eb} = 100$ Oe for CoFeB 5 nm/IrMn 23 nm/Pt 4 nm and $\theta_{eb} = 90°$ and $H_{eb} = 17$ Oe for CoFeB 5 nm/IrMn 11 nm/Pt 4 nm. We used this technique to quantify the exchange bias field for all the samples.

   Although we conducted no field cool process for any of the samples, for thicker $t_{IrMn}$ we clearly observe the unidirectional anisotropy. We suspect that the sample stage motion during the sputtering deposition may have induced the exchange bias field.

We additionally checked the existence of the exchange bias by conducting a field cool. The field cool did not alter the thickness dependence of the exchange bias shown in Fig. 4(b).

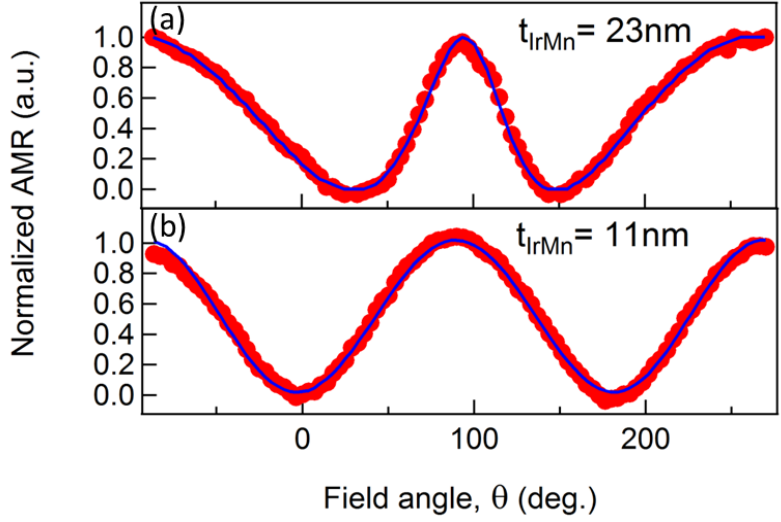

Figure S1 AMR as a function of the field angle θ for (a) CoFeB 5 nm/IrMn 23 nm/Pt 4 nm and (b) CoFeB 5 nm/IrMn 11 nm/Pt 4 nm. The blue curves are the fitting with Eq. S1.

2. Estimation of the resistivity in each layer

The conductivity of each layer is determined by measuring a set of samples varying the layer thickness of each material. The measurements are performed by a conventional two probe method. Assuming the parallel conductance model as

$$\Sigma_{tot} = (W/L) \sum_X t_X \sigma_X, \tag{S2}$$

where $X$ = CoFeB, IrMn, Pt, and $\sigma$ is the conductivity of each material. Note that we only assume the *bulk* conductance and did not take into account any *interfacial* conductance. We obtain the resistivity $1/\sigma_{CoFeB} = 1.4 \times 10^{-6}$ Ω m, $1/\sigma_{IrMn} = 1.9 \times 10^{-6}$ Ω m, and $\sigma_{Pt} = 3.0 \times 10^{-7}$ Ω m.

3. Determination of the spin Hall angle in the Pt layer

   Determination of the spin Hall angle has been a controversial issue[2,3,4,5,6,7]. Different measurement techniques give different spin Hall angle ranging from 0.01 to 0.12 for Pt. The large spread in the reported values may be due to a variation of other important parameters used to derive the spin Hall angle such as the interfacial spin mixing conductance at FM/Pt and the spin diffusion length of Pt. These parameters can be influenced significantly by the quality of the sample. In other words, how the samples are prepared can strongly affect the spin Hall angle estimation. In our experiment, we refer to the d.c. current dependence of the linewidth to estimate the spin Hall angle of the Pt layer. Fitting the data in Fig. 3(b) with

   $$\Delta\sigma/I_{dc} = \frac{\omega_0}{\gamma} \frac{\cos\theta}{(H_{ex} + 2\pi M_{eff})M_s t_{FM} w} \frac{\hbar}{2e}\left(\frac{\theta_{SH,Pt} r_{Pt}}{t_{Pt}}\right), \qquad (S3)$$

   yields $\theta_{SH,Pt} = 0.09 \pm 0.01$, which is reasonably within the range of the reported values. Note that we did not measure the spin diffusion length for the Pt or the spin mixing conductance in the sample, and that we presumed the spin diffusion length to be much smaller than the thickness and we ignored the spin mixing conductance for our spin Hall angle estimation. Therefore, we do not intend to claim a definite accuracy of the *intrinsic* spin Hall angle of Pt in this work.

4. The spin Hall angle in the IrMn layer

   It would be possible that IrMn have a spin Hall effect as Ir can give rise to the spin Hall effect [8]. We preformed ST-FMR measurements on CoFeB/IrMn/SiO$_2$ multilayers to check the possibility of IrMn spin Hall effect. If the IrMn layer gives rise to the spin Hall effect, the injected spin current into the CoFeB layer decreases or increases the linewidth of CoFeB. We tested two different IrMn thicknesses: 10 nm and 20 nm. As plotted in Fig. S2, we did not observe a clear d.c. current dependence on the linewidth. Using $1/\sigma_{CoFeB} = 1.4 \times 10^{-6}$ Ω m, $1/\sigma_{IrMn} = 1.9 \times 10^{-6}$ Ω m, and presumably $\theta_{SH,IrMn} \sim 0.05$, we estimated the expected slope $\Delta\sigma/I_{dc}$ as shown in Fig. S2. Our measurements are indeed not sensitive enough to determine spin Hall angle in the IrMn layer with two decimal places because of its high resistivity. In the discussion in the main text, we disregard the IrMn spin Hall effect for calculating $\Delta\sigma/I_{dc}$ shown in Fig. 4(a) considering the uncertainty of the value. The calculated line in Fig. 4 deviates by at most 9% even if the spin Hall angle as much as ~0.05 for IrMn is taken into account because of the small shunt current ratio in the IrMn layer.

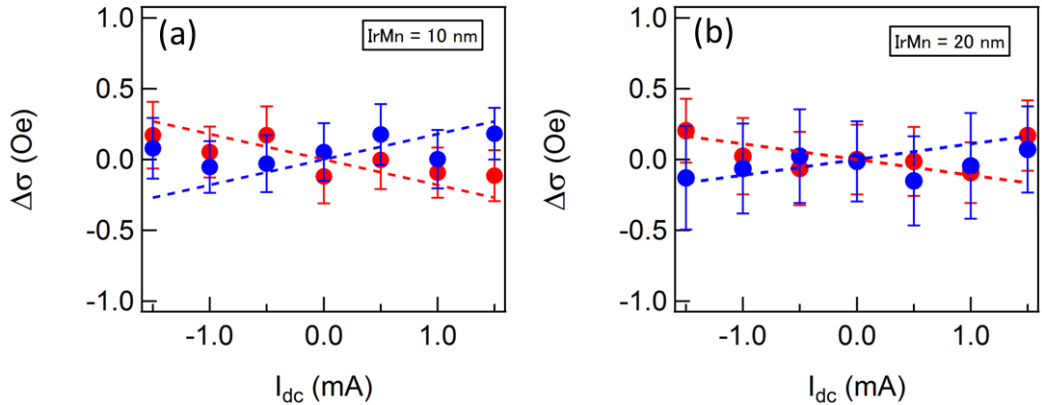

Figure S2 $\Delta\sigma$ as a function of $I_{dc}$ for (a) CoFeB 5 nm/IrMn 10 nm/SiO$_2$ 5 nm and (b) CoFeB 5 nm/IrMn 20 nm/SiO$_2$ 5 nm. The red and blue circles are respectively for the positive field and negative field. The dotted lines are estimated slope $\Delta\sigma/I_{dc}$ with $1/\sigma_{CoFeB} = 1.4 \times 10^{-6}$ Ω m, $1/\sigma_{IrMn} = 1.9 \times 10^{-6}$ Ω m, and presumably $\theta_{SH,IrMn} \sim 0.05$.